\documentclass[letterpaper,titlepage,11pt]{article}
\usepackage{latexsym,amssymb,amstext,amsmath}
\usepackage{slashed}
\usepackage{mathrsfs}
\usepackage{color}
\usepackage{enumerate}
\usepackage{graphicx}
\usepackage{tikz}
\usepackage{etex}
\usepackage{latexsym}
\usepackage{cite}
\usepackage{accents}

\allowdisplaybreaks

\usepackage[
colorlinks=true,
linkcolor=blue,
urlcolor=red,
filecolor=green,
citecolor=red,
pdfstartview=FitV,
pdftitle={},
pdfauthor={Mehmet Ozkan},
pdfsubject={},
pdfkeywords={},
pdfpagemode=None,
bookmarksopen=true
]{hyperref}

\newcommand{\bea}{\setlength\arraycolsep{2pt} \begin{eqnarray}}
\newcommand{\eea}{\end{eqnarray}}
\newcommand{\nn}{\nonumber}

\usepackage{hyperref}

\newsavebox{\uuunit}
\sbox{\uuunit}
{\setlength{\unitlength}{0.825em}
	\begin{picture}(0.6,0.7)
	\thinlines
	\put(0,0){\line(1,0){0.5}}
	\put(0.15,0){\line(0,1){0.7}}
	\put(0.35,0){\line(0,1){0.8}}
	\multiput(0.3,0.8)(-0.04,-0.02){12}{\rule{0.5pt}{0.5pt}}
	\end {picture}}

\def\be{\begin{equation}}
\def\ee{\end{equation}}
\def\ba{\begin{array}}
\def\ea{\end{array}}
\def\bea{\begin{eqnarray}}
\def\eea{\end{eqnarray}}
\def\bd{\begin{displaymath}}
\def\ed{\end{displaymath}}

\def\nn{\nonumber}

\def\a{\alpha}
\def\b{\beta}
\def\g{\gamma}

\def\d{\delta}

\def\e{\epsilon}

\def\l{\lambda}

\def\m{\mu}
\def\n{\nu}
\def\r{\rho}

\def\t{\tau}

\def\o{\omega}
\def\O{\Omega}

\def\nn{\nonumber}

\makeatletter
\@addtoreset{equation}{section}
\makeatother

\begin{document}
%
\begin{titlepage}

\bigskip

\begin{center}
{\LARGE\bfseries Three-Dimensional Higher-Order Schr\"odinger Algebras and Lie Algebra Expansions }
\\[10mm]
\textbf{Oguzhan Kasikci$^1$, Nese Ozdemir$^1$, Mehmet Ozkan$^1$ and  Utku Zorba$^{1,2}$}\\[5mm]
\vskip 25pt

{\em  \hskip -.1truecm $^1$Department of Physics, Istanbul Technical University,  \\
	Maslak 34469 Istanbul, Turkey  \vskip 10pt }

{\em  \hskip -.1truecm $^2$Van Swinderen Institute, University of Groningen,  \\
	Nijenborgh 4, 9747 AG Groningen, The Netherlands  \vskip 10pt }

{email: {\tt kasikcio@itu.edu.tr, nozdemir@itu.edu.tr, ozkanmehm@itu.edu.tr, zorba@itu.edu.tr}}

\end{center}

\vspace{3ex}

\begin{center}
{\bfseries Abstract}
\end{center}
\begin{quotation} \noindent

We provide a  Lie algebra expansion procedure to construct three-dimensional higher-order Schr\"odinger algebras which relies on a particular subalgebra of the four-dimensional relativistic conformal algebra. In particular, we reproduce the extended Schr\"odinger algebra and provide a new higher-order Schr\"odinger algebra. The structure of this new algebra leads to a discussion on the uniqueness of the higher-order non-relativistic algebras. Especially, we show that the recent d-dimensional symmetry algebra of an action principle for Newtonian gravity is not uniquely defined but can accommodate three discrete parameters. For a particular choice of these parameters, the Bargmann algebra becomes a subalgebra of that extended algebra which allows one to introduce a mass current in a Bargmann-invariant sense to the extended theory. 

\end{quotation}

\vfill

\flushleft{\today}
\end{titlepage}
\setcounter{page}{1}
\tableofcontents

\newpage

\section{Introduction}{\label{Intro}}
\paragraph{}

In recent years, it has become clear that the structure of non-relativistic gravity theories is quite different than the structure of general relativity. In particular, there are many different non-relativistic gravity theories that are based on the extensions of the Bargmann algebra that cannot be accessed from the Poincar\'e algebra by means of an Inonu-Wigner contraction. A well-known example of this kind, called the three-dimensional extended Bargmann gravity \cite{EBG}, has been studied from various angles including its supersymmetric completion and matter couplings \cite{EBGBergshoeff}, its cosmological and non-relativistic conformal  \cite{EBGHartong} and Maxwellian \cite{Concha1} extensions as well as the construction of its parity-odd cousin, the exotic extended Bargmann supergravity \cite{Ozdemir1}. Another important example that goes beyond the Bargmann symmetries, which was put forward in \cite{Bleeken,Hartong1}, is based on the argument that although the Bargmann symmetries are sufficient to establish the Poisson's equation for Newtonian gravity in a covariant manner, an action principle for Newtonian gravity requires extended symmetries beyond the standard Bargmann symmetries. It was furthermore shown in \cite{Ozdemir2} that a non-trivial central extension of the algebra of \cite{Hartong1} is necessary for three-dimensions to construct a three-dimensional Chern-Simons gravity action which was later generalized to include a cosmological constant in \cite{Concha2}. 

 All these recent progress begs for a systematic understanding of extended symmetries that goes beyond the Bargmann symmetry. There are various proposals to generate such extended algebras. From a contraction viewpoint, the non-relativistic limit of Poincar\'e $\oplus$ U(1) algebra gives rise to the Bargmann algebra. Therefore, as the number of generators increases in the non-relativistic side, one has to go beyond the Poincar\'e algebra. To this end, it was shown in \cite{CoAdjoint} by three-dimensional examples that the non-relativistic limit of coadjoint Poincar\'e algebras is a promising candidate to generate higher-order non-relativistic algebras. Similarly, in \cite{Ozdemir2}, the non-relativistic limit of a particular bi-metric theory was shown to generate an extended three-dimensional gravity, known as the extended Newtonian gravity. Instead of contraction of relativistic algebras, which preserves the number of generators, one can consider the Lie algebra expansion which generates larger Lie algebras starting from a specific one \cite{LAE1,LAE2,LAE3}. This methodology is based on a consistent truncation of infinite series expansion for Maurer-Cartan one-forms and has recently been applied extensively to non-relativistic settings to provide extended non-relativistic (super)algebras and corresponding (super)gravity models, see e.g. \cite{Ozdemir1, BergshoeffOrtin, LAEE1, LAEE2}.  

There is an important class of non-relativistic algebras that admit non-relativistic conformal symmetry. These algebras are known to provide a different scaling dimension for space and time coordinates, i.e. under scaling transformations the time coordinate scale as $t \rightarrow  \l^2 t$ while space coordinates scale as $x \rightarrow  \l x$. Therefore the structure of non-relativistic conformal algebras is quite distinct from their relativistic cousin. As a price to pay,  non-relativistic conformal algebras do not contain the spatial part of the special conformal symmetry but only include dilatations $(D)$, temporal special conformal symmetry generator $(K)$ and possible higher-order non-relativistic conformal symmetry generators. The simplest representative of such algebras is known as the Schr\"odinger algebra whose generators are the symmetries of the Schr\"odinger equation. This algebra is actually a subalgebra of the conformal group itself  \cite{sorba} and is known to play a crucial role in the condensed matter applications of non-relativistic symmetries \cite{Son}. In three dimensions, a higher-order Schr\"odinger  algebra, known as the extended Schr\"odinger algebra, was established in \cite{EBGHartong} as a non-relativistic conformal extension of the extended Bargmann algebra. Although the above-mentioned mechanisms have been widely used to generate various extensions of the Bargmann algebra, there is no systematic procedure known to generate extensions of Schr\"odinger algebra. The main purpose of this paper is to fill this gap in three-dimensions by utilizing the Lie algebra expansion methodology. In order to achieve our goal, we first introduce a particular subalgebra of three-dimensional conformal algebra that is unique to three-dimensions in Section \ref{Section2}. It is a new, novel non-relativistic conformal extension of the Galilei algebra, and as far as we know, this subalgebra has been overlooked so far. Next, we employ the Lie algebra expansion procedure and first reproduce the extended Schr\"odinger algebra of \cite{EBGHartong}, then go to the next order and establish what we call the  ``enhanced Schr\"odinger algebra" and the corresponding ``enhanced Schr\"odinger gravity".

As the extended Schr\"odinger algebra of \cite{EBGHartong} is the non-relativistic conformal extension of the extended Bargmann algebra, the construction of the higher-order enhanced Schr\"odinger algebra leads to a natural question: What algebra does one obtain by truncating the non-relativistic conformal generators from the enhanced Schr\"odinger algebra? To address this question, it is natural to see what algebra arises after the extended Bargmann algebra in the Lie algebra expansion of the three-dimensional Poincar\'e algebra. This was already achieved in \cite{BergshoeffOrtin} and it was shown that the next order algebra is the extended Newtonian algebra \cite{Ozdemir2}. To our surprise, we found that the consistent truncation of the enhanced Schr\"odinger algebra does not give rise to the extended Newtonian algebra but another algebra with the same set of generators. Our result leads us to question the uniqueness of the higher-order extended algebras and we provide several examples on the non-uniqueness at a particular order in three and higher dimensions. In Section \ref{Section4} we give conclusion and discussions.

\section{The Core Algebra of Higher Order Schr\"odinger Algebras} \label{Section2}

As mentioned, our primary goal in this paper is to create machinery to generate higher-order Schr\"odinger algebras  based on Lie algebra expansions. As this procedure is based on an infinite series expansion of a core Lie algebra with certain properties, we dedicate this section to a very brief review of  Lie algebra expansion methodology as well as the core algebra of the higher-order Schr\"odinger algebra. 

\subsection{Lie Algebra Expansions}
Lie algebra expansion is a three-step procedure that generates higher-order Lie algebras starting from a core algebra. 
\begin{enumerate}
\item {The core algebra must be a direct sum of two subspaces $g = V_0 \oplus V_1$ where $V_0$ and $V_1$ satisfy
	\bea
	\left[V_0, V_0\right] \subset V_0\,, \qquad \left[V_0, V_1\right] \subset V_1\,, \qquad \left[V_1, V_1\right] \subset V_0\,.
	\label{EvenOdd}
	\eea
	In the first step, we split the generators of the core algebra into the even $(V_0)$ and odd $(V_1)$ class of generators, i.e. if $X_i$ with $i = 1, \ldots, {\rm{dim}} g$ represents the generators of the core algebra, then we may split the generators as $X_i = ( X_{i_0}, X_{i_1})$ where $X_{i_0} \in V_0$ with $i_0 = 1\ldots {\rm{dim}}V_0$ and $X_{i_1} \in V_1$ with $i_1 = 1\ldots {\rm{dim}}V_1$.}
\item{In the second step, we assign a gauge field to each generator
	\bea
	\o^i = \left(\o^{i_0}, \o^{i_1}\right) \,,
	\label{SplitGenerator}
	\eea
	and establish their Maurer-Cartan equations based on the structure constants of the core algebra.
	\bea
	d \o^k = - \frac12 C_{ij}{}^k \o^i \wedge \o^j \,.
	\label{MaurerCartan}
	\eea
	We expand the gauge fields  in powers of an expansion parameter $\lambda$ depending on which class they belong to, i.e. gauge fields that are associated with even generators are expanded in even powers of $\lambda$ whereas the odd ones are expanded in odd powers of $\lambda$
	\bea
	\o^{i_0}  =  \sum_{\alpha_{0}=0, \, \alpha_{0}\,\, \textrm{even}}^{\infty}
	\accentset{(\alpha_{0})}{\omega}{}^{\, \, \, i_{0}}
	\lambda^{\alpha_{0}} \,, \qquad \o^{i_1} = 
	\sum_{\alpha_{1}=1, \, \alpha_{1}\,\, \textrm{odd}}^{\infty}
	\accentset{(\alpha_{1})}{\omega}{}^{\, \, \,i_{1}} \lambda^{\alpha_{1}} \,,
	\label{InfExpandGenerator}
	\eea
	where $\a_0$ and $\a_1$ indicates the order of expansion. The resulting power series are also used to expand the relevant Maurer-Cartan equations
	\bea
	d \accentset{(\gamma_{s})}{\omega}{}^{\, \,\,k_{s}}
	=
	- \frac{1}{2} C_{i_{p},\alpha_{p}\, j_{q},\beta_{q}}{}^{k_{s},\gamma_{s}}\,\,
	\accentset{(\alpha_{p})}{\omega}{}^{\,\,\, i_{p}}\wedge
	\accentset{(\beta_{q})}{\omega}{}^{\,\,\, j_{q}} \,.
	\label{ExpandMC}
	\eea
	where $s,p,q = 0,1$ and $C_{i_{p},\alpha_{p}\, j_{q},\beta_{q}}{}^{k_{s},\gamma_{s}} = C_{i_p j_q}{}^{k_s}$ if $\g_s = \a_p + \b_q$ and it vanishes otherwise.}
\item{In the final step, we truncate the infinite-dimensional algebra to a finite-dimensional one by considering the following consistent truncation
	\bea
	\o^{i_0}  =  \sum_{\alpha_{0}=0, \, \alpha_{0}\,\, \textrm{even}}^{N_0}
	\accentset{(\alpha_{0})}{\omega}{}^{\, \, \, i_{0}}
	\lambda^{\alpha_{0}} \,, \qquad \o^{i_1} = 
	\sum_{\alpha_{1}=1, \, \alpha_{1}\,\, \textrm{odd}}^{N_1}
	\accentset{(\alpha_{1})}{\omega}{}^{\, \, \,i_{1}} \lambda^{\alpha_{1}} \,,
	\label{FiniteExpandGen}
	\eea
	where the consistency is imposed by either setting $N_0 = N_1 + 1$ or $N_1 = N_0 + 1$.} As in \cite{BergshoeffOrtin} we will represent the algebras corresponding to these two conditions as $g(N_0, N_1) = g(N+1, N)$ and $g(N_0, N_1) = g(N, N+1)$ respectively. 
\end{enumerate}
Before proceeding to the actual construction of extended Schr\"odinger algebras, some comments are in order. As evident from the expansion procedure that we outlined here, higher-order algebras that we generate with Lie algebra expansion inherits the symmetry properties of core algebra. Consequently,  if a core algebra has a certain symmetry amongst its generators, this will also show itself in some form in the higher-order algebras.  Furthermore, the structure of the expanded Maurer-Cartan equation (\ref{ExpandMC}) indicates that any structure constant $C_{ij}{}^k$ with $i\neq j$ of the core algebra generates at least two different commutation relations with the same structure constants in the higher-order algebra. Thus, if a higher-order algebra does not have this property, then either the gauge fields of the corresponding generators are the zeroth-order gauge fields in the expansion (\ref{FiniteExpandGen}) or the higher-order algebra does not originate from a Lie algebra expansion. With these notes in mind, we are now ready to proceed to the construction of a core algebra that can generate higher-order Schr\"odinger algebras.

\subsection{The Core Algebra}

The implementation of the Lie algebra expansion methodology to generate higher-order Schr\"odinger algebras requires a core algebra with the essential property (\ref{EvenOdd}). We may form such an algebra by considering the four-dimensional conformal algebra
\begin{align}
\left[M_{AB}, M_{CD}\right] = & \eta_{AC} M_{BD} - \eta_{BC} M_{AD} - \eta_{AD} M_{BC} + \eta_{BD} M_{AC}  \,, &\left[M_{AB}, \widehat P_C\right] = &   \eta_{AC} \widehat P_{B} - \eta_{BC} \widehat P_{A}\,, \nn\\
\left[P_A,\widehat K_B\right] = & - 2 \left(\eta_{AB} \widehat D  +  M_{AB}\right) \,,& \left[M_{AB}, \widehat K_C\right] = &    \eta_{AC} \widehat K_{B} - \eta_{BC} \widehat K_{A} \,,\nn\\
\left[\widehat D, \widehat P_A\right] = & - \widehat P_A \,,  &\left[\widehat D, \widehat K_A\right] = & \widehat K_A \,,
\label{ConformalAlgebra}
\end{align}
where $A,B = 0, \ldots 3$. Here we have $\widehat P_A$ for translations, $M_{AB}$ for Lorentz transformations, $\widehat D$ for dilatations and $\widehat K_A$ for special conformal symmetry. Based on these commutation relations, the generators of the four-dimensional conformal algebra can be split into even $(V_0)$ and odd $(V_1)$ class of generators as
\bea
V_0 = \{\widehat P_0, \widehat P_3, \widehat K_0, \widehat K_3, \widehat D, M_{03} , M_{ab}\}\,, \qquad V_1 = \{\widehat P_a, \widehat K_a, M_{a0}, M_{a3}\} \,.
\label{EvenOddConformal}
\eea
Using the components of these generators, we may make the following definitions
\begin{align}
h =& \frac{1}{2} \left(\widehat P_0  + \widehat P_3\right) \,, & j =& \frac{1}{6} \left(\widehat P_0 - \widehat P_3  \right) + \frac{1}{6} \left(\widehat K_0 + \widehat K_3  \right) - \frac{2}{3} \widehat J \,,  & k = &\frac{1}{2} \left(\widehat K_0  - \widehat K_3\right)  \,,\nn\\
p_a =& \frac12 \left(\widehat P_a + M_{a3}^\star + M_{a0}^\star\right) \,,  & g_a =& \frac12 \left(\widehat K_a + M_{a3}^\star - M_{a0}^\star\right) \,,  & d =& \widehat D + M_{03}\,, 
\label{CoreGen}
\end{align}
where $a, b = 1,2$ and $\widehat{J} = \frac12 \e^{ab} M_{ab}$ and the starred quantities are defined as
\bea
M^\star_{a3} = \e_a{}^b M_{b0}\,, \qquad M_{a0}^\star = \e_a{}^b M_{b 3} \,.
\eea 
Note that due to the structure of the duality relation that we used here, these definitions are unique to four dimensions. The generator (\ref{CoreGen}) form a closed subalgebra of the conformal algebra that is given by
\begin{align}
\left[j, p_{a}\right] &= - \e_{ab} p^b \,, & 
\left[j, g_{a}\right]    &= - \e_{ab} g^b \,,&   
\left[h, g_a\right] &=  - \e_{ab} p^b \,, \nn\\
\left[p_a,p_b\right] &=  \e_{ab} h \,,   &  \left[g_a, g_b\right]  &= \e_{ab} k \,,  &   \left[p_a,g_b \right]  &=  -\frac12 \d_{ab} d + \frac32\e_{ab} j \,,\nn\\
\left[k, p_a\right] &= - \e_{ab} g^b \,, & \left[h, k\right] & = d\,, & \left[d, p_a\right] & = - p_a \,, \nn\\
\left[d, g_a \right] & =  g_a \,, &  \left[d, h\right] &= - 2h \,,  & \left[d, k\right] & = 2 k  \,.
\label{CoreAlgebra}
\end{align}
This subalgebra has several notable features. First of all, due to the splitting of the conformal generators (\ref{EvenOddConformal}), the algebra (\ref{CoreAlgebra}) can be written as the direct sum of an even and an odd class of generators $V_0 \oplus V_1$ where
\bea
V_0 = \{h, j, k, d\}\,, \qquad V_1 = \{p_a, g_a \} \,,
\eea
which is the first step necessary to implement the Lie algebra expansion. Second, it is invariant under the following redefinition
\bea
p_a \rightarrow g_a \,,\qquad g_a \rightarrow p_a\,, \qquad  h \rightarrow k\,, \qquad k \rightarrow h\,, \qquad j \rightarrow j \,, \qquad d \rightarrow -d \,,
\eea
which will present itself in different forms in the higher-order algebras. Third, if these generators are treated as the generators of a non-relativistic algebra, i.e. $j$ as the generators of spatial rotations, $p_a$ for spatial translations, $g_a$ for Galilean boosts, $h$ for time translations, $d$ for dilatations and $k$ for non-relativistic special conformal transformations, then  (\ref{CoreGen})  can be considered as a novel non-relativistic conformal Galilean algebra. Finally, the core algebra can be equipped with an invariant bi-linear form that we can use to construct a Chern-Simons action
\bea
(p_a,g_b)= \delta_{a b}\,, \quad (j,j) = -\frac{2}{3}\,, \quad (d,d) = 2\,,\quad (h,k)= -1\,, \label{metric}
\eea
We may construct an action principle for the core algebra using the invariant bi-linear form and the structure constants via Chern-Simons action formula
\bea
S &=& \frac{k}{4\pi} \int{\rm{Tr}} \left( A \wedge dA + \frac23 A \wedge A \wedge A \right) \,,
\eea
where $A = A_\m dx^\m$.  In order to identify $A_\m$  in terms of the generators and  the gauge  fields of the core algebra, we  assign a  gauge field to each of the generators 
\bea
p_a \rightarrow E_\m{}^a \,, \qquad g_a \rightarrow \O_\m{}^a \,, \qquad h \rightarrow \theta_\m \,, \qquad k \rightarrow F_\m \,, \qquad d \rightarrow B_\m \,, \qquad j \rightarrow \O_\m \,.
\label{CoreGauge}
\eea
Consequently,  $A_\m$ reads
\bea
A_\m = E_\m{}^a p_a + \O_\m{}^a g_a +  \theta_\m h + F_\m k + B_\m d + \O_\m j \,,
\eea
and the Chern-Simons Lagrangian for the core algebra is given by
\bea
S &=& \frac{k}{4\pi} \int d^3 x\,  \e^{\m\n\r} \Big(  E_\m{}^a R_{\n\r a}(g) + \O_\m{}^a R_{\n\r a} (p) - \theta_\m R_{\n\r}(k) - F_\m R_{\n\r} (h) + 4 B_\m \partial_\n B_\r    \nn\\
&& \qquad \qquad \qquad\quad - \frac43 \O_\m \partial_\n \O_\r   + \e_{ab} \O_\m{}^a \O_\n{}^b \theta_\r + \e_{ab} E_\m{}^a E_\n{}^b F_\r  \Big) \,. 
\label{CoreAction}
\eea
Note that it is not possible to express the core action completely in terms of the curvatures due to the non-vanishing diagonal couplings $(d,d)$ and $(j,j)$. The curvatures that we used here are defined as
\bea
R_{\m\n} (h) &=& 2 \partial_{[\m} \theta_{\n]} + \e_{ab} E_{[\m}{}^a  E_{\n]}{}^b - 4 B_{[\m} \theta_{\n]}\,,\nn\\
R_{\m\n} (j) &=& 2 \partial_{[\m} \O_{\n]} + 3 \e_{ab} E_{[\m}{}^a \O_{\n]}{}^b\,,\nn\\
R_{\m\n} (k) &=& 2 \partial_{[\m} F_{\n]} + \e_{ab} \O_{[\m}{}^a \O_{\n]}{}^b + 4 B_{[\m} F_{\n]} \,,\nn\\
R_{\m\n} (d) &=& 2 \partial_{[\m} B_{\n]} - E_{[\m}{}^a \O_{\n] a} +  2 \theta_{[\m} F_{\n]}\,,\nn\\
R_{\m\n}{}^a (p) &=& 2 \partial_{[\m} E_{\n]}{}^a  + 2 \e^{ab} \O_{[\m} E_{\n] b} + 2 \e^{ab} \theta_{[\m} \Omega_{\n] b} - 2 B_{[\m} E_{\n]}{}^a\,,\nn\\
R_{\m\n}{}^a (g) &=& 2 \partial_{[\m} \Omega_{\n]}{}^a + 2 \e^{ab} \O_{[\m} \Omega_{\n] b} + 2 \e^{ab} F_{[\m} E_{\n] b} + 2 B_{[\m} \Omega_{\n]}{}^a  \,. \label{coreCurv}
\eea

\section{Higher Order Schr\"odinger Algebras} \label{Section3}

Armed with core algebra (\ref{CoreGen}),  its properties and its Chern-Simons action (\ref{CoreAction}), we may now proceed to the second step of the Lie algebra expansion and construct higher-order non-relativistic Schr\"odinger algebras and their corresponding actions. 

\subsection{Extended Schr\"odinger Gravity}
In this subsection we will show that the $(N_0, N_1) = (2,1)$ algebra that we obtain from the Lie algebra expansion of the core algebra (\ref{CoreAlgebra}) leads to the three-dimensional extended Schr\"odinger algebra constructed in \cite{EBGHartong}.  In principle, one can start the expansion at $(1,0)$ order, however, as can be seen from (\ref{CoreAlgebra}), this would simply lead to a truncation of the Schr\"odinger algebra with the mass generator $M$. The $(1,0)$ order algebra also coincides with an Inonu-Wigner contraction of the core algebra (\ref{CoreAlgebra}) once the generators of the Galilean transformations and the spatial translations  are properly rescaled with the speed of light $c$, i.e. $g_a \rightarrow c g_a$ and $p_a \rightarrow c p_a$. As the $g(1,0)$ algebra do not admit a non-degenerate invariant bi-linear form, we skip this step and start with $g(2,1)$ case. To proceed to the second step of the Lie algebra expansion procedure, we first establish the curvatures that correspond to the gauge fields of the core algebra (\ref{CoreGauge}). Imposing the consistent truncation $(N_0, N_1) = (2,1)$ the expansion of the gauge fields can be given by
\begin{align}
\theta_\m =& \accentset{(0)}{\theta}_\m + \l^2 \accentset{(2)}{\theta}_\m \,, & \O_\m =& \accentset{(0)}{\O}_\m + \l^2 \accentset{(2)}{\O}_\m \,, & F_\m = & \accentset{(0)}{F}_\m + \l^2 \accentset{(2)}{F}_\m \,,\nn\\
B =&  \accentset{(0)}{B}_\m + \l^2 \accentset{(2)}{B}_\m \,, & E_\m{}^a = & \lambda \accentset{(1)}{E}_\m{}^a\,, & \O^a = & \lambda \accentset{(1)}{\O}_\m{}^a \,.
\end{align}
To finalize the second step of the Lie algebra expansion methodology, we use the expansion of the gauge fields expand the curvatures as 
\bea
\accentset{(0)}{R}_{\m\n} (h) &=& 2 \partial_{[\m} \accentset{(0)}{\theta}_{\n]} - 4 \accentset{(0)}{B}_{[\m} \accentset{(0)}{\theta}_{\n]}\,,\nn\\
\accentset{(2)}{R}_{\m\n} (h) &=& 2 \partial_{[\m} \accentset{(2)}{\theta}_{\n]} + \e_{ab} \accentset{(1)}{E}_{[\m}{}^a  \accentset{(1)}{E}_{\n]}{}^b - 4 \accentset{(0)}{B}_{[\m} \accentset{(2)}{\theta}_{\n]} - 4 \accentset{(2)}{B}_{[\m} \accentset{(0)}{\theta}_{\n]}  \,,\nn\\
\accentset{(0)}{R}_{\m\n} (j) &=& 2 \partial_{[\m} \accentset{(0)}{\O}_{\n]} \,,\nn\\
\accentset{(2)}{R}_{\m\n} (j) &=& 2 \partial_{[\m} \accentset{(2)}{\O}_{\n]} + 3 \e_{ab} \accentset{(1)}{E}_{[\m}{}^a \accentset{(1)}{\O}_{\n]}{}^b\,,\nn\\
\accentset{(0)}{R}_{\m\n} (k) &=& 2 \partial_{[\m} \accentset{(0)}{F}_{\n]} + 4 \accentset{(0)}{B}_{[\m} \accentset{(0)}{F}_{\n]} \,,\nn\\
\accentset{(2)}{R}_{\m\n} (k) &=& 2 \partial_{[\m} \accentset{(2)}{F}_{\n]} + \e_{ab} \accentset{(1)}{\O}_{[\m}{}^a  \accentset{(1)}{\O}_{\n]}{}^b + 4 \accentset{(0)}{B}_{[\m} \accentset{(2)}{F}_{\n]} + 4 \accentset{(2)}{B}_{[\m} \accentset{(0)}{F}_{\n]}  \,,\nn\\
\accentset{(0)}{R}_{\m\n} (d) &=& 2 \partial_{[\m} \accentset{(0)}{B}_{\n]} + 2 \accentset{(0)}{\theta}_{[\m} \accentset{(0)}{F}_{\n]} \,,\nn\\
\accentset{(2)}{R}_{\m\n} (d) &=& 2 \partial_{[\m} \accentset{(2)}{B}_{\n]} + \delta_{ab} \accentset{(1)}{\O}_{[\m}{}^a  \accentset{(1)}{E}_{\n]}{}^b + 2 \accentset{(0)}{\theta}_{[\m} \accentset{(2)}{F}_{\n]} + 2 \accentset{(2)}{\theta}_{[\m} \accentset{(0)}{F}_{\n]}  \,,\nn\\
\accentset{(1)}{R}_{\m\n}{}^a (p) &=& 2 \partial_{[\m} \accentset{(1)}{E}_{\n]}{}^a  + 2 \e^{ab} \accentset{(0)}{\O}_{[\m} \accentset{(1)}{E}_{\n] b} + 2 \e^{ab} \accentset{(0)}{\theta}_{[\m} \accentset{(1)}{\O}_{\n] b} - 2 \accentset{(0)}{B}_{[\m} \accentset{(1)}{E}_{\n]}{}^a\,,\nn\\
\accentset{(1)}{R}_{\m\n}{}^a (g) &=& 2 \partial_{[\m} \accentset{(1)}{\O}_{\n]}{}^a + 2 \e^{ab} \accentset{(0)}{\O}_{[\m} \accentset{(1)}{\O}_{\n] b}  + 2 \e^{ab} \accentset{(0)}{F}_{[\m} \accentset{(1)}{E}_{\n] b} + 2 \accentset{(0)}{B}_{[\m} \accentset{(1)}{\O}_{\n]}{}^a  \,.
\label{ExpandedCurvaturesSch}
\eea
We can proceed to the third step and read off the structure constants and construct the $g(2,1)$ algebra 
\begin{align}
\left[ H, G_a \right]  &= - \e_{ab} P^b\,,          &  \left[ J, P_a \right] &= - \e_{ab} P^b\,,          &  \left[ J, G_a \right] &=- \e_{ab} G^b\,, \nonumber\\
\left[ P_{a}, G_b \right]  &= \e_{ab} M + \d_{ab} Y \,,        &  \left[ G_a, G_{b} \right]  &= \e_{ab} S \,, & \left[ P_{a}, P_b \right]  &= \e_{ab} Z \,,  \nn\\
\left[ D, G_a \right]  &= G_{a} \,,          &  \left[ D, P_a \right] &= -  P_{a} \,,          &  \left[ D, H \right] &=- 2 H\,, \nonumber\\
\left[ D, S \right]  &=2 S \,,        &  \left[ D,K \right]  &= 2K \,, & \left[ D,Z\right]  &= - 2 Z \,, \nn\\
\left[ K, P_{a} \right]  &= - \e_{ab} G^{b} \,,          &  \left[ K, H\right] &= -  D \,,          &  \left[ K, Y \right] &= S \,, \nonumber\\
\left[ K, Z \right]  &=2 Y \,,        &  \left[ H, S \right]  &= - 2 Y \,, & \left[ H, Y \right]  &= - Z \,, 
\label{SchBoson}
\end{align}
where we  identify the generators as
\begin{align}
\accentset{(0)}{h} = & H \,, & \accentset{(2)}{h} = & Z\,, & \accentset{(0)}{j} =& J \,,  & \accentset{(2)}{j} =& \frac23 M \,,  & \accentset{(0)}{k} =& K\,,  \nn\\
\accentset{(2)}{k} =& S\,, &\accentset{(0)}{d} = & D \,, & \accentset{(2)}{d} = & - 2 Y \,, & \accentset{(1)}{p_a} =& P_a\,,  & \accentset{(1)}{g_a} =&  G_a \,,
\label{SchGenIden}
\end{align} 
This algebra is precisely the extended Schr\"odinger algebra \cite{EBGHartong}. Note that due to the structure of its core algebra, the extended Schr\"odinger algebra is invariant under following redefinitions
\begin{align}
P_a \rightarrow& \, G_a \,,& G_a \rightarrow& \, P_a\,, &  H \rightarrow& \, K\,, & K \rightarrow& \, H\,, & J \rightarrow& \, J \,,\nn\\
Z \rightarrow& \, S\,,& S\rightarrow&\, Z \,, &  M \rightarrow& \, M\,, & Y \rightarrow& - Y\,.
\end{align}
We may also expand the Chern-Simons Lagrangian for the core algebra (\ref{CoreAction}), which is given by
\bea
S &=& \frac{k}{4\pi} \int d^3 x\,  \e^{\m\n\r}  \Big( e_\m{}^a R_{\n\r a}(G) + \o_\m{}^a R_{\n\r a} (P) - 2 s_\m R_{\n\r}(H) - 2 m_\m R_{\n\r}(J) \nn\\
&& \qquad  \qquad \qquad \quad -2 y_\m R_{\n\r}(D) - 2 z_\m R_{\n\r}(K) \Big)\,,
\label{ESG}
\eea
where we relabelled the gauge fields as
\begin{align}
\accentset{(0)}{\theta}_\m = & \tau_\m \,, & \accentset{(2)}{\theta}_\m = & z_\m\,, & \accentset{(0)}{\O}_\m =& \o_\m\,,  & \accentset{(2)}{\O}_\m =& \frac32 m_\m \,,  & \accentset{(0)}{F}_\m =& f_\m \,,  \nn\\
\accentset{(2)}{F}_\m =& s_\m \,, &\accentset{(0)}{B}_\m = & b_\m \,, & \accentset{(2)}{B}_\m = & - \frac12 y_\m \,, & \accentset{(1)}{E}_\m{}^a =& e_\m{}^a\,,  & \accentset{(1)}{\o}_\m{}^a =&  \o_\m{}^a \,,
\label{SchFieldIden}
\end{align} 
and the curvatures can simply be read off from (\ref{ExpandedCurvaturesSch}) with the identifications (\ref{SchGenIden}) and (\ref{SchFieldIden})
\bea
R_{\m\n}(H) &=& 2 \partial_{[\m} \t_{\n]}  - 4 b_{[\m} \t_{\n]}  \,,\nn\\
R_{\m\n}{}^a (P) &=&  2 \partial_{[\m} e_{\n]}{}^a + 2 \e^{ab}\,  \o_{[\m} e_{\n]b} - 2 \e^{ab}\,  \o_{ [\m b} \t_{\n]}  - 2 b_{[\m} e_{\n]}^a \,,\nn\\
R_{\m\n}(M) &=& 2 \partial_{[\m} m_{\n]} + 2 \e_{ab} \o_{[\m}{}^a e_{\n]}{}^b  \,,\nn\\
R_{\m\n}{}^a (G) &=& 2 \partial_{[\m} \o_{\n]}{}^a + 2  \epsilon{}^{ab} \o_{[\m}{} \o_{\n] b}  + 2  \epsilon{}^{ab} f_{[\m}{} e_{\n] b} + 2 b_{[\m} \o_{\n]}^a  \,,\nn\\
R_{\m\n} (J) &=& 2 \partial_{[\m} \o_{\n]}\,,\nn\\
R_{\m\n} (S) &=& 2 \partial_{[\m} s_{\n]}  +  \epsilon{}^{ab} \o_{[\m a} \o_{\n]b }   +  4 b_{[\m} s _{\n]} + 2 f_{[\m} y_{\n]} \,,\nn\\
R_{\m\n} (K) &=& 2 \partial_{[\m} f_{\n]}   +  4 b_{[\m} f _{\n]}  \,,\nn\\
R_{\m\n} (D) &=& 2 \partial_{[\m} b_{\n]}   +  2 \t_{[\m} f _{\n]}  \,,\nn\\
R_{\m\n} (Y) &=& 2 \partial_{[\m} y_{\n]}  -  2  \o _{[\m a} e_{\n] }^a +  4 f_{[\m}  z_{\n]} - 4 \t_{[\m} s_{\n]}  \,,\nn\\
R_{\m\n} (Z) &=& 2 \partial_{[\m} z_{\n]}  +  \epsilon{}^{ab} e_{[\m a} e_{\n]b }  - 4  b_{[\m} z_{\n]} - 2 \t_{[\m} y _{\n]} \,.
\eea
The action (\ref{ESG}) is precisely the extended Schr\"odinger gravity of \cite{EBGHartong}.
\subsection{Enhanced Schr\"odinger Gravity}

In this subsection, we study the case $g(4,3)$. In principle, this should produce the Schr\"odinger extension of the extended Newtonian algebra \cite{Ozdemir2} as it is the next order algebra after the extended Bargmann algebra in the Lie algebra expansion of the three-dimensional Poincar\'e algebra \cite{BergshoeffOrtin}. However, as we will see, it produces a Schr\"odinger extension of a different algebra with the same generators as the extended Newtonian algebra.  This result leads us to question the uniqueness of the $g(4,3)$ algebras and as we will discuss in Subsection \ref{Section33}, such algebras are not unique. Deferring this discussion for the next subsection, we consider  the same setting as in the previous subsection but now we truncate the expansion at $g(4,3)$ order
\begin{align}
\theta_\m =& \accentset{(0)}{\theta}_\m + \l^2 \accentset{(2)}{\theta}_\m + \l^4 \accentset{(4)}{\theta}_\m \,, & \O_\m =& \accentset{(0)}{\O}_\m + \l^2 \accentset{(2)}{\O}_\m +  \l^4 \accentset{(4)}{\O}_\m \,, & F_\m = & \accentset{(0)}{F}_\m + \l^2 \accentset{(2)}{F}_\m +  \l^4 \accentset{(4)}{F}_\m \,,\nn\\
B =&  \accentset{(0)}{B}_\m + \l^2 \accentset{(2)}{B}_\m  + \l^4 \accentset{(4)}{B}_\m\,, & E_\m{}^a = & \lambda \accentset{(1)}{E}_\m{}^a  + \lambda^3 \accentset{(3)}{E}_\m{}^a\,, & \O^a = & \lambda \accentset{(1)}{\O}_\m{}^a +  \lambda^3 \accentset{(3)}{\O}_\m{}^a \,. \label{gaugeg(3,4)}
\end{align}
Having the expansion of the gauge fields, we obtain the corresponding curvatures via (\ref{coreCurv}) as 
\bea
\accentset{(0)}{R}_{\m\n} (h) &=& 2 \partial_{[\m} \accentset{(0)}{\theta}_{\n]} - 4 \accentset{(0)}{B}_{[\m} \accentset{(0)}{\theta}_{\n]}\,,\nn\\
\accentset{(2)}{R}_{\m\n} (h) &=& 2 \partial_{[\m} \accentset{(2)}{\theta}_{\n]} + \e_{ab} \accentset{(1)}{E}_{[\m}{}^a  \accentset{(1)}{E}_{\n]}{}^b - 4 \accentset{(0)}{B}_{[\m} \accentset{(2)}{\theta}_{\n]} - 4 \accentset{(2)}{B}_{[\m} \accentset{(0)}{\theta}_{\n]}  \,,\nn\\
\accentset{(4)}{R}_{\m\n} (h) &=& 2 \partial_{[\m} \accentset{(4)}{\theta}_{\n]} + 2\e_{ab} \accentset{(1)}{E}_{[\m}{}^a  \accentset{(3)}{E}_{\n]}{}^b - 4 \accentset{(0)}{B}_{[\m} \accentset{(4)}{\theta}_{\n]} - 4 \accentset{(4)}{B}_{[\m} \accentset{(0)}{\theta}_{\n]} - 4 \accentset{(2)}{B}_{[\m} \accentset{(2)}{\theta}_{\n]}  \,,\nn\\
\accentset{(0)}{R}_{\m\n} (j) &=& 2 \partial_{[\m} \accentset{(0)}{\O}_{\n]} \,,\nn\\
\accentset{(2)}{R}_{\m\n} (j) &=& 2 \partial_{[\m} \accentset{(2)}{\O}_{\n]} + 3 \e_{ab} \accentset{(1)}{E}_{[\m}{}^a \accentset{(1)}{\O}_{\n]}{}^b\,,\nn\\
\accentset{(4)}{R}_{\m\n} (j) &=& 2 \partial_{[\m} \accentset{(4)}{\O}_{\n]} + 3 \e_{ab} \accentset{(1)}{E}_{[\m}{}^a \accentset{(3)}{\O}_{\n]}{}^b + 3 \e_{ab} \accentset{(3)}{E}_{[\m}{}^a \accentset{(1)}{\O}_{\n]}{}^b\,,\nn\\
\accentset{(0)}{R}_{\m\n} (k) &=& 2 \partial_{[\m} \accentset{(0)}{F}_{\n]} + 4 \accentset{(0)}{B}_{[\m} \accentset{(0)}{F}_{\n]} \,,\nn\\
\accentset{(2)}{R}_{\m\n} (k) &=& 2 \partial_{[\m} \accentset{(2)}{F}_{\n]} + \e_{ab} \accentset{(1)}{\O}_{[\m}{}^a  \accentset{(1)}{\O}_{\n]}{}^b + 4 \accentset{(0)}{B}_{[\m} \accentset{(2)}{F}_{\n]} + 4 \accentset{(2)}{B}_{[\m} \accentset{(0)}{F}_{\n]}  \,,\nn\\
\accentset{(4)}{R}_{\m\n} (k) &=& 2 \partial_{[\m} \accentset{(4)}{F}_{\n]} + 2 \e_{ab} \accentset{(1)}{\O}_{[\m}{}^a  \accentset{(3)}{\O}_{\n]}{}^b + 4 \accentset{(0)}{B}_{[\m} \accentset{(4)}{F}_{\n]}  + 4 \accentset{(4)}{B}_{[\m} \accentset{(0)}{F}_{\n]}+ 4 \accentset{(2)}{B}_{[\m} \accentset{(2)}{F}_{\n]}  \,,\nn\\
\accentset{(0)}{R}_{\m\n} (d) &=& 2 \partial_{[\m} \accentset{(0)}{B}_{\n]} + 2 \accentset{(0)}{\theta}_{[\m} \accentset{(0)}{F}_{\n]} \,,\nn\\
\accentset{(2)}{R}_{\m\n} (d) &=& 2 \partial_{[\m} \accentset{(2)}{B}_{\n]} + \delta_{ab} \accentset{(1)}{\O}_{[\m}{}^a  \accentset{(1)}{E}_{\n]}{}^b + 2 \accentset{(0)}{\theta}_{[\m} \accentset{(2)}{F}_{\n]} + 2 \accentset{(2)}{\theta}_{[\m} \accentset{(0)}{F}_{\n]}  \,,\nn\\
\accentset{(4)}{R}_{\m\n} (d) &=& 2 \partial_{[\m} \accentset{(4)}{B}_{\n]} + \delta_{ab} \accentset{(3)}{\O}_{[\m}{}^a  \accentset{(1)}{E}_{\n]}{}^b +  \delta_{ab} \accentset{(1)}{\O}_{[\m}{}^a  \accentset{(3)}{E}_{\n]}{}^b+ 2 \accentset{(0)}{\theta}_{[\m} \accentset{(4)}{F}_{\n]} + 2 \accentset{(4)}{\theta}_{[\m} \accentset{(0)}{F}_{\n]} + 2 \accentset{(2)}{\theta}_{[\m} \accentset{(2)}{F}_{\n]}\,,\nn\\
\accentset{(1)}{R}_{\m\n}{}^a (p) &=& 2 \partial_{[\m} \accentset{(1)}{E}_{\n]}{}^a  + 2 \e^{ab} \accentset{(0)}{\O}_{[\m} \accentset{(1)}{E}_{\n] b} + 2 \e^{ab} \accentset{(0)}{\theta}_{[\m} \accentset{(1)}{\O}_{\n] b} - 2 \accentset{(0)}{B}_{[\m} \accentset{(1)}{E}_{\n]}{}^a\,,\nn\\
\accentset{(3)}{R}_{\m\n}{}^a (p) &=& 2 \partial_{[\m} \accentset{(3)}{E}_{\n]}{}^a  + 2 \e^{ab} \accentset{(0)}{\O}_{[\m} \accentset{(3)}{E}_{\n] b} + 2 \e^{ab} \accentset{(2)}{\O}_{[\m} \accentset{(1)}{E}_{\n] b}+ 2 \e^{ab} \accentset{(0)}{\theta}_{[\m} \accentset{(3)}{\O}_{\n] b} + 2 \e^{ab} \accentset{(2)}{\theta}_{[\m} \accentset{(1)}{\O}_{\n] b}\nn\\
&&  - 2 \accentset{(0)}{B}_{[\m} \accentset{(3)}{E}_{\n]}{}^a - 2 \accentset{(2)}{B}_{[\m} \accentset{(1)}{E}_{\n]}{}^a\,,\nn\\
\accentset{(1)}{R}_{\m\n}{}^a (g) &=& 2 \partial_{[\m} \accentset{(1)}{\O}_{\n]}{}^a + 2 \e^{ab} \accentset{(0)}{\O}_{[\m} \accentset{(1)}{\O}_{\n] b}  + 2 \e^{ab} \accentset{(0)}{F}_{[\m} \accentset{(1)}{E}_{\n] b} + 2 \accentset{(0)}{B}_{[\m} \accentset{(1)}{\O}_{\n]}{}^a \,,\nn \\
\accentset{(3)}{R}_{\m\n}{}^a (g) &=& 2 \partial_{[\m} \accentset{(3)}{\O}_{\n]}{}^a + 2 \e^{ab} \accentset{(0)}{\O}_{[\m} \accentset{(3)}{\O}_{\n] b}  + 2 \e^{ab} \accentset{(2)}{\O}_{[\m} \accentset{(1)}{\O}_{\n] b}+ 2 \e^{ab} \accentset{(0)}{F}_{[\m} \accentset{(3)}{E}_{\n] b} +  2 \e^{ab} \accentset{(2)}{F}_{[\m} \accentset{(1)}{E}_{\n] b} \nn\\
&& + 2 \accentset{(0)}{B}_{[\m} \accentset{(3)}{\O}_{\n]}{}^a + 2 \accentset{(2)}{B}_{[\m} \accentset{(1)}{\O}_{\n]}{}^a \,,
\label{EnhancesedCurvaturesSch}
\eea 
Once again, we can read off the structure constants from the expanded curvatures and construct the $g(4,3)$ algebra 
\begin{align}
[D,H] & = -2 H  &  \quad  [D,Z] &= -2 Z  &\quad  [D,K] & = 2 K \nn\\
[D,S] & = 2 S  &  [H,K] &=D &  [H,S] &= -2Y \nn \\
[H,Y] & = -Z  &   [K,Y] &=S &  [K,Z] &= 2Y\nn \\ 
[H,G_{a}]&= - \epsilon_{a b} P^{b} & [H,B_{a}] & = -\epsilon_{a b} T^{b} & [Z,G_{a}] & = - \epsilon_{a b} T^{b} \nn \\
[Y,G_{a}]&=-\frac{1}{2} B_{a} & [J,G_{a}] & =- \epsilon_{a b} G^{b} & [J,B_{a}] & = - \epsilon_{a b} B^{b} \nn  \\
[M,G_{a}]&= - \frac{3}{2} \epsilon_{a b} B^{b} & [J,P_{a}] & = - \epsilon_{a b} P^{b} &  [J,T_{a}] & = - \epsilon_{a b} T^{b}\nn  \\
[M,P_{a}]&=- \frac{3}{2} \epsilon_{a b} T^{b} &   
[K,P_{a}]&=- \epsilon_{a b} G^{b} &
[S,P_{a}]&=- \epsilon_{a b} B^{b}   \nn\\
[K,T_{a}]&=- \epsilon_{a b} B^{b}&  
[G_{a},G_{b}]&=\epsilon_{a b} S  &  
[G_{a},P_{b}]&=- \delta_{a b} Y+  \epsilon_{a b}   M \nn  \\
[P_{a},P_{b}]&=\epsilon_{a b} Z &
[D,P_{a}]&=-P_{a} &  
[D,T_{a}]&=-T_{a} \nn  \\
[Y,P_{a}]&=\frac{1}{2} T_{a}&  
[D,G_{a}]&=G_{a} &  
[D,B_{a}]&=B_{a} \nn \\ 
[P_{a},T_{b}]&=\epsilon_{a b} X_1 &
[G_{a},B_{b}]&=\epsilon_{a b} X_3 &
[P_{a},B_{b}]&= \delta_{a b} X_4+  \epsilon_{a b}   X_2 \nn \\
[T_{a},G_{b}]&= \delta_{a b} X_4+  \epsilon_{a b}   X_2 &
[D,X_1] & = -2 X_1 &
[Y,Z] & =  X_1 \nn \\
[X_4,H] & =  X_1 &
[D,X_3] & = 2 X_3  &
[S,Y] & =  X_3  \nn \\
[K,X_4] & =  X_3 &
[X_1,K] & = -2 X_4 & 
[Z,S] & =  -2 X_4  \nn \\
[H,X_3] & = -2 X_4. 
\label{EnhancedSch}
\end{align}
where we identify the generators as 
\begin{align}
\accentset{(4)}{h} = & X_1 \,, & \accentset{(4)}{j} = & \frac{2}{3} X_2\,, & \accentset{(4)}{k} =& X_3 \,, \nn \\
\accentset{(4)}{d} =& - 2 X_4 \,, & \accentset{(3)}{p_a} =& T_a\,,  & \accentset{(3)}{g_a} =&  B_a \,,
\label{EnhancedSchGenIden}
\end{align} 
together with our previous definitions (\ref{SchGenIden}). Similarly, the gauge fields identified as 
\begin{align}
\accentset{(4)}{\theta}_\m = & x_{1\m} \,, & \accentset{(4)}{\Omega}_\m = & \frac{3}{2}x_{2\m} \,, & \accentset{(4)}{F}_\m =& x_{3\m} \,,  \nn\\
 \accentset{(4)}{B}_\m =& - \frac12  x_{4\m}  \,,  
& \accentset{(3)}{E}_\m{}^a =& t_\m{}^a\,,  & \accentset{(3)}{\O}_\m{}^a =&  b_\m{}^a \,.
\label{SchFieldIdeng(3,4)}
\end{align} 
 together with our previous definitions (\ref{SchFieldIden}).  We refer the algebra (\ref{EnhancedSch}) as the enhanced Schr\"odinger algebra. The enhanced Schr\"odinger algebra can be equipped with an invariant bi-linear form to form a Chern-Simons action
 \begin{align}
 &(P_a,B_b)= \delta_{a b}, & \quad 
 &(T_a,G_b)= \delta_{a b}, & \quad  
 &(J,X_2) =-1, \nn \\
 &(M,M)= -\frac32, & \quad 
 &(D,X_4)=-1, & \quad 
 &(Y,Y)=\frac12,   \nn\\
 &(H,X_3)=-1, & \quad 
 &(K,X_1)=-1, & \quad 
 &(S,Z)=-1 \,.
 \end{align}
Alternatively, one can consider the Lie algebra expansion of the core action (\ref{CoreAction}), which gives rise to the enhanced Schr\"odinger gravity
\bea
S &=& \frac{k}{4\pi} \int d^3 x\,   \e^{\m\n\r}  \Big( b_\m{}^a R_{\n\r a}(P)  + e_\m{}^a R_{\n\r a}(B) + \o_\m{}^a R_{\n\r a} (T) + t_\m{}^a R_{\n\r a} (G) - x_{1\m} R_{\n\r}(K)  \nn\\
&& \qquad \qquad \qquad \quad - f_\m R_{\n\r}  (X_1)  - z_\m R_{\n\r}(S) - s_\m R_{\n\r}(Z) - \t_\m R_{\n\r}(X_3) - x_{3\m} R_{\n\r}(H) \nn\\
&&  \qquad   \qquad \qquad \quad + y_\m \partial_\n y_\r -3 m_\m \partial_\n m_\r - 4 x_{4\m} \partial_\n b_\r - 4 x_{2\m} \partial_\n \o_\r   + \e_{ab} \, \o_\m{}^a \o_\n{}^b z_\r    \nn\\
&&  \qquad   \qquad \qquad \quad  + 2 \e_{ab}\, \o_\m{}^a b_\n{}^b \t_\r   + \e_{ab}\,  e_\m{}^a e_\n{}^b s_\r   + 2 \e_{ab} \, e_\m{}^a t_\n{}^b f_\r    \Big)\,,
\eea
Here, the curvatures can simply be read off from (\ref{EnhancedSch}) with the identifications  (\ref{EnhancedSchGenIden}) and (\ref{SchFieldIdeng(3,4)})  
\bea
R_{\m\n}{}^a (P) &=&  2 \partial_{[\m} e_{\n]}{}^a + 2 \e^{ab}\,  \o_{[\m} e_{\n]b} - 2 \e^{ab}\,  \o_{ [\m b} \t_{\n]}  - 2 b_{[\m} e_{\n]}^a \,,\nn\\
R_{\m\n}{}^a (G) &=& 2 \partial_{[\m} \o_{\n]}{}^a + 2  \epsilon{}^{ab} \o_{[\m}{} \o_{\n] b}  + 2  \epsilon{}^{ab} f_{[\m}{} e_{\n] b} + 2 b_{[\m} \o_{\n]}^a  \,,\nn\\
R_{\m\n}{}^a  (B) &=&  2 \partial_{[\m} b_{\n]}{}^a  + \e^{ab} \o_{[\m} b_{\n]}{}^b + 3 \e^{ab} m_{[\m} \o_{\n]b} + 2 \e^{ab} f_{[\m} t_{\n]b} + 2\e^{ab} s_{[\m} e_{\n] b}\nn\\
&& + 2 b_{[\m} b_{\n]}{}^a - y_{[\m} \o_{\n]}{}^a  \,,\nn\\
R_{\m\n}{}^a (T) &=&  2 \partial_{[\m} t_{\n]}{}^a   + \e^{ab} \o_{[\m} t_{\n]}{}^b + 3 \e^{ab} m_{[\m}  e_{\n]b} + 2 \e^{ab} \t_{[\m} b_{\n]b} + 2\e^{ab} z_{[\m} \o_{\n] b}\nn\\
&& - 2 b_{[\m} t_{\n]}{}^a + y_{[\m} e_{\n]}{}^a \,,\nn\\
R_{\m\n}(H) &=& 2 \partial_{[\m} \t_{\n]}  - 4 b_{[\m} \t_{\n]}  \,,\nn\\
R_{\m\n} (K) &=& 2 \partial_{[\m} f_{\n]}   +  4 b_{[\m} f _{\n]}  \,,\nn\\
R_{\m\n} (S) &=& 2 \partial_{[\m} s_{\n]}  +  \epsilon{}^{ab} \o_{[\m a} \o_{\n]b }   +  4 b_{[\m} s _{\n]} + 2 f_{[\m} y_{\n]} \,,\nn\\
R_{\m\n} (Z) &=& 2 \partial_{[\m} z_{\n]}  +  \epsilon{}^{ab} e_{[\m a} e_{\n]b }  - 4  b_{[\m} z_{\n]} - 2 \t_{[\m} y _{\n]} \,,\nn\\
R_{\m\n}(X_1) &=&  2 \partial_{[\m} x_{1\n]} + 2 \e_{ab} e_{[\m}{}^a t_{\n]}{}^b - 4 b_{[\m} x_{1\n]} + 2 x_{4[\m} \t_{\n]} + 2 y_{[\m} z_{\n]} \,,\nn\\
R_{\m\n}(X_2) &=&  2 \partial_{[\m} x_{2\n]} + 2 \e_{ab} e_{[\m}{}^a b_{\n]}{}^b +  2 \e_{ab} t_{[\m}{}^a \o_{\n]}{}^b \,,\nn\\
R_{\m\n}(X_3) &=& 2 \partial_{[\m} x_{3\n]} + 2 \e_{ab} \o_{[\m}{}^a b_{\n]}{}^b + 4 b_{[\m} x_{3\n]} + 2 x_{4[\m} f_{\n]} - 2 y_{[\m} s_{\n]} \,\nn \\
R_{\m\n}(X_4) &=& 2 \partial_{[\m} x_{4\n]} + 2 \delta_{ab} e_{[\m}{}^a b_{\n]}{}^b + 2 \delta_{ab} t_{[\m}{}^a \o_{\n]}{}^b - 4 x_{1[\m} f_{\n]} - 4 \t_{[\m} s_{\n]} + 4 x_{3[\m} \t_{\n]} \,.
\eea

\subsection{Non-Uniqueness of  Higher Order Algebras}\label{Section33}

In the previous subsection, we obtained a $g(4,3)$ Schr\"odinger algebra and the corresponding $g(4,3)$ Schr\"odinger gravity. Following this result, there is an obvious question to address: What algebra does one obtain by truncating the non-relativistic conformal generators from the $g(4,3)$ Schr\"odinger algebra?  To answer this question, we  may truncate the enhanced Schr\"odinger algebra by  consistently truncating $\{D, K, X_1, X_2, X_3, X_4\}$ in which case one obtains
\begin{align}
[H,S] & = -2Y\,, &[H,Y]  &= -Z  \,,  & [H,G_{a}]  & = - \epsilon_{a b} P^{b}\,, \nn \\
[H,B_{a}]  &= -\epsilon_{a b} T^{b} \,, &[Z,G_{a}]  & = - \epsilon_{a b} T^{b} \,, &[Y,G_{a}] &=-\frac{1}{2} B_{a}\,, \nn \\ 
[J,G_{a}]  &=- \epsilon_{a b} G^{b}  \,, &[J,B_{a}]  & = - \epsilon_{a b} B^{b} \,, &[M,G_{a}] &= - \frac{3}{2} \epsilon_{a b} B^{b} \,, \nn \\
[J,P_{a}]  &= - \epsilon_{a b} P^{b} \,, &[J,T_{a}]  &= - \epsilon_{a b} T^{b} \,, &[M,P_{a}] &=- \frac{3}{2} \epsilon_{a b} T^{b} \,,\nn \\   
[S,P_{a}] & =- \epsilon_{a b} B^{b} \,,  &  [G_{a},G_{b}] & =\epsilon_{a b} S  \,,  & [G_{a},P_{b}] & =- \delta_{a b} Y +  \epsilon_{a b}   M\,,  \nn \\
[P_{a},P_{b}] &=\epsilon_{a b} Z \,,  &[Y,P_{a}] & =\frac{1}{2} T_{a} \,.
\label{New}
\end{align}
Although they have the same generators, the algebra (\ref{New}) is different than the three-dimensional extended Newtonian algebra that is given by \cite{Ozdemir2}
\begin{align}
\left[H, G_a\right] & = P_a \,, & [J,P_{a}]  &= - \epsilon_{a b} P^{b} \,, &[J,G_{a}]  &=- \epsilon_{a b} G^{b}  \,, \nn\\
[J,T_{a}]  &= - \epsilon_{a b} T^{b} \,, & [J,B_{a}]  & = - \epsilon_{a b} B^{b} \,, & [G_{a},P_{b}] & = \epsilon_{a b}   M \,,\nn\\
[G_{a},G_{b}] & =\epsilon_{a b} S  \,,  &[H,B_{a}]  &= -\epsilon_{a b} T^{b} \,, & [M,G_{a}] &=- \epsilon_{a b} T^{b} \,,\nn\\
[S,P_{a}] & =- \epsilon_{a b} T^{b} \,, & [S,G_{a}] & =- \epsilon_{a b} B^{b} \,, &[G_{a},T_{b}] & = - \epsilon_{a b} Y \,, \nn\\
[G_{a},B_{b}] & =\epsilon_{a b} Z \,, & [P_{a},B_{b}] & = - \epsilon_{a b} Y \,.
\label{ENA}
\end{align}
These two algebras most importantly differ by the fact that while $Y$ and $Z$ are central in the extended Newtonian algebra (\ref{ENA}), they are not central and cannot be consistently truncated in (\ref{New}). This result is somewhat puzzling and it leads us to question how unique the higher-order algebras are. In the rest of this section, we will discuss several examples of non-uniqueness in three and higher dimensions.

\subsubsection*{Example 1:} Consider the following algebra with four parameters $a_1, a_2,a_3$ and $a_4$
	\begin{align}
	\left[J, G_a\right] & = - \e_{ab} G^b \,, & \left[J, P_a\right] & = - \e_{ab} P^b \,,  &\left[J, B_a\right] & = - \e_{ab} B^b \,, \nn\\
	 \left[J, T_a\right] & = - \e_{ab} T^b \,,  &	\left[H, B_a\right] & = - \e_{ab} T^b \,, & \left[H, G_{a}\right]& = -\e_{ab} P^b \,, \nn\\
	  \left[G_a, G_b\right] & = \e_{ab} S \,, & \left[M, G_a\right] & = a_1 \e_{ab} T^b  \,, &\left[S, G_a\right] & =  a_1 \e_{ab} B^b \,, \nn\\
	    \left[S, P_a\right] & =  a_1 \e_{ab} T^b \,, &\left[B_a, G_b\right] & =  a_2 \e_{ab} Z \,, &  \left[B_a, P_b\right] & = a_3 \e_{ab} Y  \,, \nn\\
	  \left[G_a, P_b\right] & =  \frac12 a_4 \d_{ab} Y + \e_{ab} M\,,&	\left[G_a, T_b\right] & =  a_3 \e_{ab} Y \,, & \left[H,S\right] & = a_4 Y \,.
	\end{align}
If all the parameters are non-zero, they can be absorbed into the redefinitions of the generators. However, we may construct discrete set of  algebras by setting certain parameters to zero. Of particular interest, for the following choice of the parameters 
\bea
a_1 = -1\,, \quad  a_2 = 1\,, \quad a_3 = -1\,, \quad  a_4 = 0\,,
\eea
 this algebra reduces to the three-dimensional  extended Newtonian algebra \cite{Ozdemir2}. We may lift this algebra to arbitary dimensions by setting $a_2 = a_3 = a_4 = 0$ and thereby truncating $Y$ and $Z$ as
	\begin{align}
 \left[J_{ab}, G_c\right] & = 2 \delta_{c[b} G_{a]} \,, &  \left[J_{ab}, G_c\right] & = 2 \delta_{c[b} P_{a]}  \,,  & \left[J_{ab}, B_c\right] & = 2 \delta_{c[b} B_{a]} \,, \nn\\
   \left[J_{ab}, T_c\right] & = 2 \delta_{c[b} T_{a]} \,,  & \left[H, B_a\right] & = T_{a}\,, & \left[H, G_{a}\right]& = P_{a} \,, \nn\\
    \left[G_a, G_b\right] & = S_{ab} \,, & \left[M, G_a\right] & = - a_1 T_{a}  & \left[S_{ab}, G_c\right] & =  -2 a_1 \delta_{c[b} B_{a]} \,, \nn\\
      \left[S_{ab}, P_c\right] & =   - 2 a_1 \delta_{c[b} T_{a]} \,, & \left[G_a, P_b\right] & =  \delta_{ab} M\,, & \left[J_{ab}, J_{cd}\right] & = 4 \delta_{[a[c}J_{d]b]} \,,\nn\\
      \left[J_{ab}, S_{cd}\right] & = 4 \delta_{[a[c}S_{d]b]} \,.
      \label{HartongObers}
	\end{align}
Note that the three-dimensional algebra can be recovered by setting 
\bea
J_{ab} \rightarrow \e_{ab} J\,, \qquad S_{ab} \rightarrow \e_{ab} S\,, \qquad P^a \rightarrow \e^{ab} P_b\,,\qquad T^a \rightarrow \e^{ab} T_b \,.
\eea
This algebra precisely match with the higher-order algebra of \cite{Hartong1}  upon rescaling
\bea
B_a \rightarrow \frac{1}{a_1} B_a \,, \qquad T_a \rightarrow \frac{1}{a_1} T_a
\eea
as long as $a_1 \neq 0$. However, if $a_1 = 0$, then we obtain an extension of the Bargmann algebra that is spanned by the generators $\{H, J_{ab}, P_{a}, G_{a}, M\}$ with four extra generators $\{S_{ab}, B_{a}, T_{a}\}$ where $M$ stays central unlike the algebra of \cite{Hartong1}.

\subsubsection*{Example 2:}

Next, we consider the following algebra with four parameters $a_1, a_2,a_3$ and $a_4$
\begin{align}
\left[J, G_a\right] & = - \e_{ab} G^b \,, & \left[J, P_a\right] & = - \e_{ab} P^b \,,  &\left[J, B_a\right] & = - \e_{ab} B^b \,, \nn\\
\left[J, T_a\right] & = - \e_{ab} T^b \,,  &	\left[H, B_a\right] & = - \e_{ab} T^b \,, & \left[H, G_{a}\right]& = -\e_{ab} P^b \,, \nn\\
\left[G_a, G_b\right] & = \e_{ab} S \,, & \left[M, G_a\right] & = a_3 \e_{ab} T^b  \,, &\left[S, G_a\right] & =  a_3 \e_{ab} B^b \,, \nn\\
\left[S, P_a\right] & =  a_3 \e_{ab} T^b \,, &\left[B_a, G_b\right] & =  a_{4} \e_{ab} Z \,, &  \left[G_a, P_b\right] & =  \frac12 a_1 \d_{ab} Y + \e_{ab} M \,, \nn\\
\left[P_a, P_b\right] &= \frac12 a_1 a_2 \e_{ab}  Z \,, & \left[H,S\right] & = a_1 Y \,, &\left[H,Y\right] &= a_2 Z \,, 
\end{align}
Once again, the closure of the algebra does not require any specific choice for the free parameters and various algebras  can be obtained by setting certain parameters to zero. Unlike the first example, this algebra does not reduce either to the extended Newtonian algebra (\ref{ENA}) or the new algebra with the same generators (\ref{New}) for any choice of the discrete parameters. Nonetheless, it can be lifted to higher dimensions by setting $a_1 = a_2 = a_4 = 0$ to obtain the algebra (\ref{HartongObers}).

\subsubsection*{Example 3:}

As a third example, we consider the following algebra with four parameters $a_1, a_2,a_3$ and $a_4$
\begin{align}
\left[J, G_a\right] & = - \e_{ab} G^b \,, & \left[J, P_a\right] & = - \e_{ab} P^b \,,  &\left[J, B_a\right] & = - \e_{ab} B^b \,, \nn\\
\left[J, T_a\right] & = - \e_{ab} T^b \,,  &	\left[H, B_a\right] & = - \e_{ab} T^b \,, & \left[H, G_{a}\right]& = -\e_{ab} P^b \,, \nn\\
\left[G_a, G_b\right] & = \e_{ab} S \,, & \left[M, G_a\right] & = a_3 \e_{ab} T^b  \,, &\left[S, G_a\right] & =  a_3 \e_{ab} B^b \,, \nn\\
\left[S, P_a\right] & =  a_3 \e_{ab} T^b \,, &\left[B_a, P_b\right] & =  a_{4} \e_{ab} Y \,, &  \left[G_a, P_b\right] & =  \frac12 a_1 \d_{ab} Y + \e_{ab} M \,, \nn\\
 \left[G_a, T_b\right] & = a_4 \e_{ab} Y\,, & \left[Z, G_a\right] & = a_2 \e_{ab} T^b \,, &\left[H,S\right] & = a_1 Y \,. 
\end{align}
The closure of the algebra, again, is independent of the choice of  parameters. Furthermore, as in the second example, this algebra does not reduce either to the extended Newtonian algebra (\ref{ENA}) or the new algebra with the same generators (\ref{New}) for any choice of the parameters. We may set different paramters to zero to investigate various subalgebras. In particular, setting $a_1 = a_2 = a_4 = 0$, we may obtain the d-dimensional algebra (\ref{HartongObers}).

\subsubsection*{Example 4:}

Consider the following algebra with four parameters $a_1, a_2,a_3$ and $a_4$
\begin{align}
\left[J, G_a\right] & = - \e_{ab} G^b \,, & \left[J, P_a\right] & = - \e_{ab} P^b \,,  &\left[J, B_a\right] & = - \e_{ab} B^b \,, \nn\\
\left[J, T_a\right] & = - \e_{ab} T^b \,,  &	\left[H, B_a\right] & = - \e_{ab} T^b \,, & \left[H, G_{a}\right]& = -\e_{ab} P^b \,, \nn\\
\left[G_a, G_b\right] & = \e_{ab} S \,, & \left[M, G_a\right] & = \frac34 a_1 a_2 a_3 \e_{ab} B^b + a_4 \e_{ab} T^b  \,,  &\left[M, P_a\right] & = \frac34 a_1 a_2 a_3 \e_{ab} T^b\nn\\
\left[S, G_a\right] & =  a_4 \e_{ab} B^b \,, & \left[S, P_a\right] & =  \frac12 a_1 a_2 a_3 \e_{ab} B^b + a_4 \e_{ab} T^b  \,, & \left[Y, G_a\right] & = -\frac12 a_2 a_3 B_a \,, \nn\\
 \left[G_a, P_b\right] & =  \frac12 a_1 \d_{ab} Y + \e_{ab} M \,, & \left[Y, P_a\right] & = \frac12 a_2 a_3 T_a \,, & \left[Z,G_a\right]& = a_3 \e_{ab} T^b \,,\nn\\
  \left[P_a, P_b\right] &= \frac12 a_1 a_2 \e_{ab} Z \,, & \left[H,S\right] & = a_1 Y \,,  &\left[H,Y\right] &= a_2 Z \,.
\end{align}
There is no constraint on the parameters for the closure of this algebra and various discrete set of different algebras by setting certain parameters to zero . In particular we may set 
\bea
a_1 = - 2\,, \quad a_2 = -1\,, \quad  a_3 = -1 \,, \quad a_4 = 0 \,,
\eea
in which case we recover the new algebra (\ref{New}). We may also choose to set all the parameters to zero in which case we obtain an extension of the Bargmann algebra with four extra generators $\{S_{ab}, B_{a}, T_{a}\}$ where $M$ stays central as we discussed in Example 1.

\subsubsection*{Example 5:}

So far, we investigated the three-dimensional models such that when it is possible to lift them to higher dimensions, they give rise to the same algebra of \cite{Hartong1} with a single parameter. In our final example, we show that the algebra of \cite{Hartong1} can actually accommodate  three parameters
\begin{align}
\left[J_{ab}, G_c\right] & = 2 \delta_{c[b} G_{a]} \,, &  \left[J_{ab}, P_c\right] & = 2 \delta_{c[b} P_{a]}  \,,  & \left[J_{ab}, B_c\right] & = 2 \delta_{c[b} B_{a]} \,, \nn\\
\left[J_{ab}, T_c\right] & = 2 \delta_{c[b} T_{a]} \,,  & \left[H, B_a\right] & = a_1 T_{a}\,, & \left[H, G_{a}\right]& = P_{a} \,, \nn\\
\left[G_a, G_b\right] & = a_2 S_{ab} \,, & \left[M, G_a\right] & = - a_1 a_2 a_3 T_{a}  & \left[S_{ab}, G_c\right] & =  2 a_3 \delta_{c[b} B_{a]} \,, \nn\\
\left[S_{ab}, P_c\right] & =    2 a_1 a_3 \delta_{c[b} T_{a]} \,, & \left[G_a, P_b\right] & =  \delta_{ab} M\,, & \left[J_{ab}, J_{cd}\right] & = 4 \delta_{[a[c}J_{d]b]} \,,\nn\\
\left[J_{ab}, S_{cd}\right] & = 4 \delta_{[a[c}S_{d]b]} \,.
\label{HartongObersFree}
\end{align}
Here we insist that the Bargmann algebra is a subalgebra so we did not place a parameter in the relevant commutators. The closure of this algebra does not depend on any particular choice of the parameters. If all the parameters are non-zero, then they can be absorbed by into the redefinitions of the generators and we obtain the algebra of \cite{Hartong1}. However, if any one of the parameters is set to zero, then $M$ becomes central. In that case, the Bargmann algebra becomes a subalgebra and one can introduce a mass current in a Bargmann-invariant sense to the extended theory. 

\section{Discussion}\label{Section4}

In this paper, we present a Lie algebra expansion methodology to generate higher-order three-dimensional Schr\"odinger algebras. Our construction relies on a new novel three-dimensional non-relativistic conformal Galilei algebra that we utilized as a core algebra. By employing the Lie algebra expansions, we first recovered the extended Schr\"odinger algebra \cite{EBGHartong} and obtained a new higher-order Schr\"odinger algebra which we referred to as the enhanced Schr\"odinger algebra. We, next, truncate the non-relativistic conformal symmetry generators and found a new algebra that goes beyond the three-dimensional extended Bargmann algebra. Although with the same set of generators, this new algebra does not coincide with the extended Newtonian algebra \cite{Ozdemir2}. This result leads to a natural question of whether higher-order algebras are unique. We, then, showed that the higher-order algebras are indeed not unique and they can accommodate parameters such that we can obtain a set of discrete algebras by setting these parameters to zero . In particular, we showed that the symmetry algebra that was proposed as the symmetry algebra of an action for Newtonian gravity \cite{Hartong1} is not uniquely defined but can be closed with three parameters. We also show that for a particular choice of these parameters the Bargmann algebra becomes a subalgebra of the extended algebra and one can introduce a mass current in a Bargmann-invariant sense to the extended theory. 

The most straightforward continuation of this work would be to find the supersymmetric completion of the core algebra (\ref{CoreAlgebra}) therefore extending the higher-order Schr\"odinger algebras with supersymmetry. As the Schr\"odinger algebra itself if a particular instance of the $\ell$-conformal Galilei algebra \cite{Negro, Henkel}, it would also be interesting to understand the higher-order Schr\"odinger algebras by understanding the $\ell$-conformal Galilei algebra from a Lie algebra expansion viewpoint. Finally, the non-uniqueness of the higher-order non-relativistic algebras certainly deserves a detailed investigation. In our examples, the parameters were discrete such that we obtain certain algebras by setting some of the paramters to zero. It is also possible to generalize the algebra of  \cite{Hartong1} with continuous set of free parameters, see Appendix C of \cite{Bleeken1}.  In particular, given the fact that the proposed symmetry algebra of an action for Newtonian gravity can accommodate three discrete (or continuous \cite{Bleeken1}) parameters, it would be interesting to understand the metric formulation of the Lagrangian proposed in \cite{Hartong1} in terms of a gauge theory of a symmetry algebra. Finally, as the extended Schr\"odinger algebra has a hidden relativistic structure \cite{Sorokin},  it would be of interest to see if the enhanced Schr\"odinger algebra and the corresponding gravity theory can also be recast in a manifestly relativistic form, perhaps with a co-adjoint Poincar\'e structure.

\section*{Acknowledgment}

We thank Eric Bergshoeff and Dieter van den Bleeken for various comments and clarifications. The work of O.K is supported by TUBITAK grant 118F091. M.O. is supported in part by TUBITAK grant 118F091. N.O. is supported in part by Istanbul Technical University Research Fund under grant number TDK-2018-41133. U.Z. is supported in part by Istanbul Technical University Research Fund under grant number TDK-2018-41133 and TUBITAK 2214-A grant 1059B14180080.

\providecommand{\href}[2]{#2}\begingroup\raggedright\endgroup

\end{document}